\documentclass[10pt,nofootinbib]{article}
\pdfoutput=1
\usepackage[english]{babel}
\usepackage{cite,amsmath,amssymb,amsbsy,amstext, amsthm, simplewick}
\usepackage{hyperref}
\usepackage{graphicx}
\usepackage{amsfonts}
\usepackage{amssymb}
\usepackage[small]{caption}
\usepackage{upgreek}
\usepackage{exscale,relsize}
\usepackage[makeroom]{cancel}
\usepackage{soul}
\usepackage{lscape}
\usepackage{indentfirst}
\usepackage{latexsym}
\usepackage{multirow}
\usepackage{tabls}
\usepackage{epsfig}
\usepackage{bm}
\usepackage{mathrsfs}
\usepackage{comment}
\usepackage{xspace}
\usepackage[usenames,dvipsnames]{color}
\allowdisplaybreaks[1]

\allowdisplaybreaks[1]


\numberwithin{equation}{section}
\def\nn{{\nonumber}}

\def\beq{\begin{equation}}
\def\eeq{\end{equation}}
\def\bea{\begin{eqnarray}}
\def\eea{\end{eqnarray}}

\def\0{{\boldsymbol 0}}

\def\bk{{\boldsymbol{k}}}
\def\bq{{\boldsymbol{q}}}
\def\bp{{\boldsymbol{p}}}

\def\bx{{\boldsymbol{x}}}

\def\bea{\begin{eqnarray}}
\def\eea{\end{eqnarray}}
\def\nn{{\nonumber}}

\def\bk{{\boldsymbol{k}}}
\def\bp{{\boldsymbol{p}}}
\def\bq{{\boldsymbol{q}}}

\def\bx{{\boldsymbol{x}}}

\def\bA{{\boldsymbol{A}}}

\def\bD{{\boldsymbol{D}}}
\def\bE{{\boldsymbol{E}}}

\def\addICTP{\small  ICTP South American Institute for Fundamental Research\\ Rua Dr. Bento Teobaldo Ferraz 271, 01140-070 S\~ao Paulo, SP Brazil}
\begin{document}

\begin{titlepage}

\setcounter{page}{1} \baselineskip=15.5pt \thispagestyle{empty}

\begin{center}

{\fontsize{20}{28}\selectfont  \sffamily  The {\it{Lamb shift}} and the gravitational binding energy\\ [0.2cm] for binary black holes}

\end{center}

\begin{center}
\fontsize{12}{16} \selectfont \sffamily Rafael A.~Porto\\
\addICTP
\end{center}

\vspace{1.2cm}
\hrule \vspace{0.3cm}
We show that the correction to the gravitational binding energy for binary black holes due to the {\it tail effect} resembles the {\it Lamb shift} in the Hydrogen atom. In~both cases a {\it conservative} effect arises from interactions with {\it radiation} modes, and moreover an explicit cancelation between near and far zone divergences is at work. In addition, regularization scheme-dependence may introduce `ambiguity parameters'. This is remediated --within an effective field theory approach-- by the implementation of the  {\it zero-bin} subtraction. We illustrate the procedure explicitly for the Lamb shift, by performing an ambiguity-free derivation within the framework of non-relativistic electrodynamics. We also derive the renormalization group equations from which we reproduce Bethe logarithm (at order $\alpha_e^5 \log \alpha_e$), and likewise the contribution to the gravitational potential from the tail effect (proportional to $v^8 \log v$). 
\vskip 10pt
\hrule

 \end{titlepage}

\section{Introduction}


Binary coalescences are posed to become standard sources for present and future gravitational wave (GW) observatories \cite{Abadie:2010cf,Abbott:2016blz,Sesana}. GW astronomy will map the contents of the universe to an unprecedented level \cite{2016htt,2016pea}, addressing fundamental problems in astrophysics and cosmology. 
The searches demand state-of-the-art numerical and analytical modeling, to enable the most precise parameter estimation \cite{blanchet,Buoreview,Tiec}. Motivated by the construction of an accurate template bank, the effective field theory (EFT) framework was introduced to solve for the gravitational dynamics of inspiralling binary systems to high level of precision \cite{nrgr,nrgrs,walter,rafric,iragrg,rafgrg,riccardo,review}. The EFT approach was originally coined Non-Relativistic General Relativity (NRGR) \cite{nrgr}, following similarities with the techniques used for the strong interaction (NRQCD), as well as electrodynamics (NRQED). NRGR has enabled the computation of all the ingredients for the GW phase for spinning compact binary systems up to third Post-Newtonian (3PN) order \cite{prl,nrgrs1s2,nrgrs1s1,nrgr2pn,nrgr3pn,nrgrso,rads1,andirad,andirad2,review}. In addition, significant progress has been achieved towards 4PN accuracy in the EFT approach, both for non-spinning \cite{nrgr4pn1,nrgr4pn2,nltail} and rotating bodies \cite{levinnloso,levinnloss}. Some of these results have been obtained using other (more traditional) methods, see e.g. \cite{blanchet,Buoreview} for references.\vskip 4pt

The gravitational binding potential for binary systems has been recently computed in the Arnowitt, Deser, and Misner (ADM) and `Fokker-action' approaches up to 4PN order for non-spinning bodies \cite{4pnjs,4pnjs2,4pndjs,4pndjs2,4pnbla,4pnbla2,4pndj}. Despite the remarkable feat, the derivation could not be completed at first, because of regularization ambiguities. Hence, the final expression was obtained after comparison with gravitational self-force calculations \cite{4pndjs,4pnbla2}, see also \cite{ALT2}. In a companion paper \cite{comp} we describe the procedure which yields the gravitational potential, in NRGR, without the need of `ambiguity parameters'. The purpose of the present paper is to demonstrate that the issue at hand is actually more common than it might seem, since similar considerations apply in electrodynamics, and in particular in the derivation of the {\it Lamb shift} \cite{Lamb,Bethe,dyson,Viki,Kroll}. As we shall see, by performing the calculation within the EFT approach NRQED, both infrared (IR) and ultraviolet (UV) divergences are present, as in the gravitational case. 
We perform the {\it zero-bin} subtraction \cite{zero} and arrive at an ambiguity-free result. We also derive the renormalization group equation for the binding potential, and readily obtain Bethe logarithm. We then show how the manipulations in electrodynamics closely resemble the computations in gravity. In particular, the renormalization group evolution and logarithmic contributions to the binding energy may be obtained in both cases without worrying about the subtleties of the matching conditions \cite{nltail}. Throughout this paper we work in  $c=\hbar=1$ units, unless otherwise noted.

\section{The (quantum) binding energy in electrodynamics}\label{lamb}

\begin{figure}[h!]
\centerline{{\includegraphics[width=0.35\textwidth]{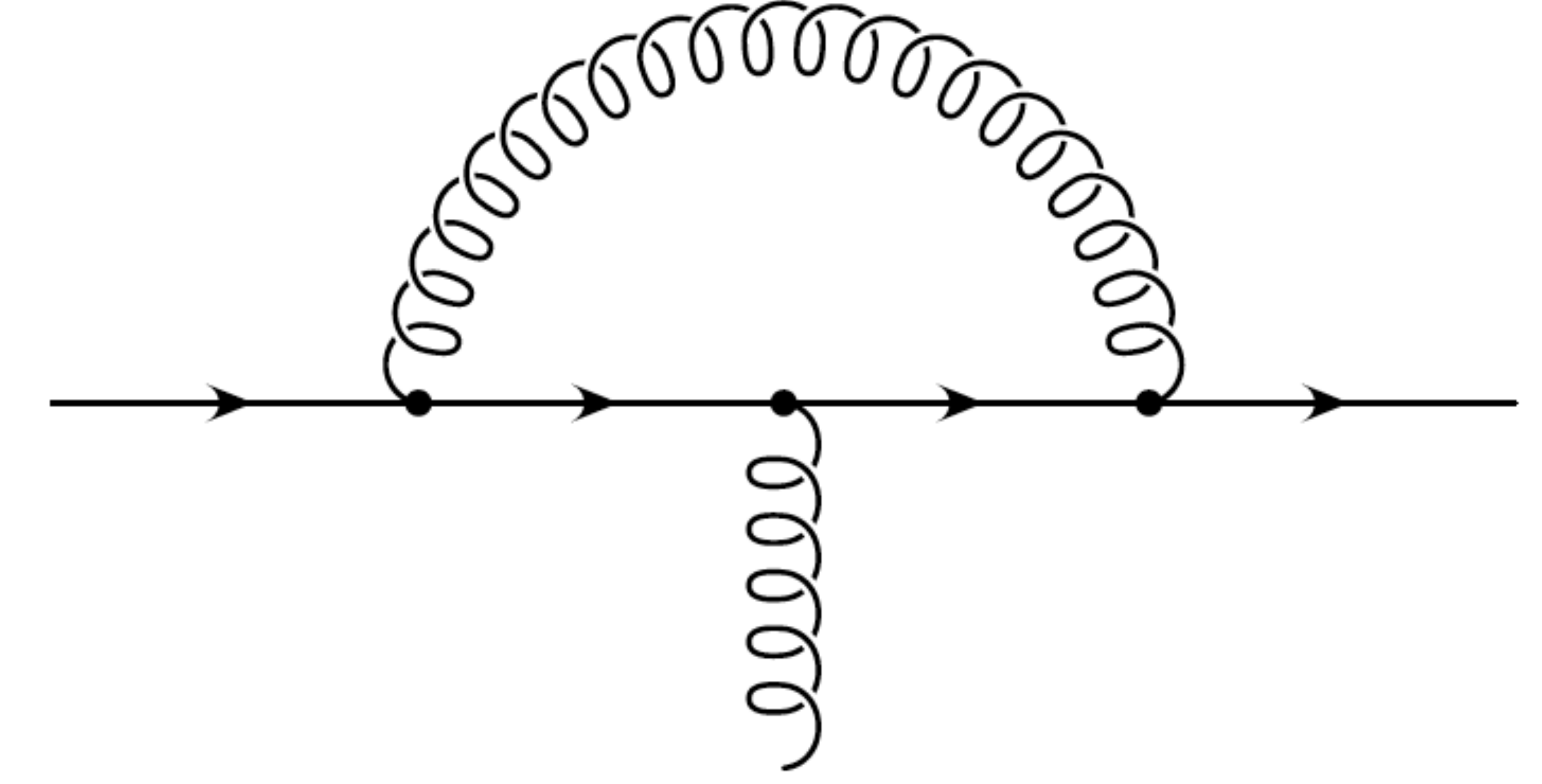}}}
\caption[1]{One loop vertex correction in electrodynamics.} \label{fig1}
\end{figure}

Quantum effects in QED contribute to the binding energy of the Hydrogen atom. A celebrated example is the Lamb shift \cite{Lamb,Bethe,dyson,Viki,Kroll}, which involves a one-loop vertex correction, see fig.~\ref{fig1}. Here we perform the computation using an EFT approach, highlighting the similarities with the binary inspiral case. We~show the existence of IR/UV divergences, discuss the zero-bin subtraction and lack of ambiguities, and the renormalization group structure.

\subsection{Form factors}
The full QED vertex (including wave-function renormalization) can be expressed in terms of two form factors,
\beq
-i e \bar u(p_1)\,\left[F_1(q^2)\gamma^\mu + \frac{i}{2m_e} F_2(q^2) \sigma^{\mu\nu} q_\nu\right] u(p_2)\,,
\eeq
with $q=p_1-p_2$, $\gamma^\mu$ the Dirac matrices, $\sigma^{\mu\nu} \equiv \frac{i}{2} [\gamma^\mu,\gamma^\nu]$, and $u(p)$ a Dirac spinor. The expressions for $F_1,F_2$ are divergent, and in dimensional regularization (dim. reg.) are given by, e.g. \cite{manohar},
\bea
F_1(q^2)&=& 1-\frac{\alpha_e(\mu)}{\pi} \frac{q^2}{m_e^2} \left[\frac{1}{3\epsilon_{\rm IR}}+\frac{1}{8} -\frac{1}{6}  \log \frac{m^2_e}{\bar\mu^2} \right] + {\cal O}(q^4)\,,\label{1}\\
F_2(q^2) &=& \frac{\alpha_e(\mu)}{2\pi} \left[1+
 \frac{q^2}{6m_e^2}\right] 
+ {\cal O}(q^4)\label{2}\,,
\eea
where $\alpha_e \equiv e^2(\mu)/4\pi$ is the fine-structure constant, $m_e$ the mass of the electron, and we have expanded to order $q^2/m_e^2$ the resulting integrals. The factor of $\bar\mu^2 \equiv 4\pi e^{-\gamma_E}\mu^2$, with $\gamma_E$ the Euler constant, appear in dim. reg. as the `subtraction scale'.\footnote{In the expressions below we omitted the bar in the $\log\bar\mu$'s, for convenience. The distinction is irrelevant for our purposes.}\vskip 4pt We will encounter both IR as well as UV divergences, which in dim. reg. emerge as poles in $\epsilon_{\rm IR/UV} \equiv (d-4)_{\rm IR/UV}$, as we approach $d=4$ dimensions. While intermedia UV divergences are present, the final expressions for the form factors are UV finite, featuring instead an IR pole (often regularized with a photon mass).\footnote{The form factor in \eqref{1} also enters in the scattering amplitude, and the IR pole is ultimately removed from the cross section by including IR divergences from (ultra-)soft photon emission \cite{weinberg}. However, as we shall see, for the binding energy the low-energy modes contribute a UV divergence instead. This is reminiscent to the gravitational scenario, where the IR divergences in the radiative multipoles turn into UV poles in the computation of the gravitational potential \cite{nltail} (see below).} 

From \eqref{1} and \eqref{2} we can derive for instance the one-loop correction to the scattering amplitude in QED, and the Lamb shift. However, in order to draw parallels with computations in gravity, in what follows we will perform the calculation within the framework of non-relativistic QED (NRQED). 

\subsection{The EFT framework: NRQED}

In addition to the electron's mass, we have two other relevant scales in the bound state problem. There is Bohr's radius, \beq r_B \simeq 1/(m_e v),\eeq with $v$ the relative velocity, and the typical frequency scale given by the Rydberg energy \beq E \simeq m_ev^2,\eeq which determines the split between levels. In a bound state the virial theorem implies \beq \alpha_e/r_B \sim m_e v^2 \to \alpha_e \sim v.\eeq 

After one eliminates the heavy scale in the theory, $m_e$, as in the heavy quark effective theory (HQET), we are left with three relevant regions \cite{Beneke,Griess,benira}: potential modes scaling as \beq (p_{\rm pot}^0,\bp_{\rm pot}) \sim (m_e v^2, m_e v) \sim (v/r_B,1/r_B)\,,\eeq 
{\it soft} modes, 
\beq
(p_S^0,\bp_S) \sim (m_e v, m_e v) \sim (1/r_B,1/r_B)\,,
\eeq 
and {\it ultra-soft} ones, \beq (p_{US}^0,\bp_{US}) \sim (m_e v^2, m_e v^2) \sim (v/r_B,v/r_B)\,.\eeq 
Notice these power counting rules are similar to the ones in NRGR, for potential and radiation fields.\footnote{The (on-shell) soft modes are not present in classical computations, since they {\it kick} the massive particle (e.g. the electron) off of the mass shell, $E \sim m_e v^2$.} The effective Lagrangian density for NRQED takes the form (ignoring spin interactions for simplicity) \cite{eftCL,manohar,benira}
\beq
{\cal L}_{\rm NRQED} = - \frac{1}{4} F^{\mu\nu} F_{\mu\nu}+ \psi^{v\dagger}_e \left(i D_0 +  \frac{\bD^2}{2m_e} +  \frac{\bD^4}{8m^3_e} + e \frac{c_V}{8m_e^2} \,   \boldsymbol{\nabla}\cdot \bE  + \cdots \right)\psi^v_e +  i \psi^{\dagger}_p  D_0 \psi_p+\cdots\,,\eeq
where $D_\mu$ is the covariant derivative, $\psi^v_e$ is given by $\psi^v_e = e^{im_e t} \psi_e$, as in HQET, and we have kept only the terms which are relevant for our purposes. We have also added the contribution from the proton, $\psi_p$, which we treat as a static source, up to ${\cal O}(m_e/m_p)$ corrections. The matching coefficient, $c_V$, is given by \cite{manohar}
\beq
c_V = F_1(0)+ 2F_2 (0)+ 8 m_e^2 \frac{d}{d q^2} F_1(0)\,,
\eeq
with the form factors in \eqref{1} and \eqref{2}. In dim. reg. the expression for $c_V$ reads
\beq
\label{eqcd}
c_V = 1 + \frac{8}{3} \frac{\alpha_e(\mu)}{\pi} \left[ - \frac{1}{\epsilon_{\rm IR}}+  \log m_e/\mu\right]\,.
\eeq
Notice we have kept the IR pole explicitly, and will be carried over until the end of the calculation. 
We will discuss later on in section \ref{zero} how to properly handle this divergence prior to computing the Lamb shift. As we shall demonstrate, this IR pole will be linked to a UV singularity arising from the ultra-soft sector. (This will be intimately related to cancelation of factors of $\log \mu$.)\vskip 4pt

The next step is to integrate out the potential and soft modes. This procedure matches NRQED into an effective theory with ultra-soft degrees of freedom only, called `potential' NRQED or pNRQED for short \cite{pnrqed}. The binding energy now becomes a matching coefficient. Therefore, we have a Coulomb-type potential of the form \cite{pnrqed},\footnote{We may construct first an EFT at the scale $m_e v$, integrating out the potential modes. In that case the interaction becomes non-local in space, but local in time \cite{IraTasi}.}
\beq
\int dt \int d^3 \bx_1 d^3\bx_2 \; \psi_p^\dagger(t,\bx_1)\psi_p(t,\bx_1) \left( \frac{\alpha_e}{|\bx_1-\bx_2|}\right) \psi_e^\dagger(t,\bx_2)\psi_e(t,\bx_2)\,.
\eeq
For the term proportional to $c_V$ we may use Gauss' law, obtaining \cite{pnrqed}
\beq
-c_V \frac{e^2}{8m_e^2} \int dt \int d^3 \bx_1 d\bx_2 \; \psi_p^\dagger(t,\bx_1)\psi_p(t,\bx_1) \psi_e^\dagger(t,\bx_2)\psi_e(t,\bx_2) \delta^3(\bx_1-\bx_2)\,.
\eeq
Since the typical size of the bound state is given by $r_B \ll 1/E$, the ultra-soft photon field is multipole expanded in powers of $E\, r_B  \sim v \sim \alpha_e$. This is reminiscent of the construction of the radiation theory in NRGR, in terms of a series of multipole moments \cite{andirad}. At the end of the day, the relevant pieces in the pNRQED Lagrangian are\footnote{ The coupling to ultra-soft photons can be re-written in a manifestly gauge invariant manner in terms of the electric field, $  \bE_{US}= -\partial_0 \bA_{US} -\boldsymbol{\nabla}_i \bA_{US}^0$, leading to a traditional dipole-type interaction: $e\,\bx\cdot \bE_{US}$. However, the expression in \eqref{pnrqed} leads to a more transparent derivation of the Lamb shift in Coulomb gauge, since the $\bA^{0}_{US}$ is a (non-propagating) constrained variable in this gauge.}\cite{manohar,pnrqed}
\bea
L_{\rm pNRQED} &=& \int d^3\bx\, \psi^\dagger(t,\bx)\left( i\partial_0 - eA^0_{US}(t,0) + e\, \bx^i \boldsymbol{\nabla}_i  A^0_{US}(t,0) + \frac{\boldsymbol{\nabla}^2}{2m_e} - V(\bx) \right.\label{pnrqed} \\ && \left. - ie \frac{\bA_{US}(t,0) \cdot \boldsymbol{\nabla}}{m_e} - c_V \frac{e^2}{8m_e} \delta^3(\bx) \right)\psi(t,\bx) - \frac{1}{4} \int d^3\bx \, F_{US}^{\mu\nu}F_{US\,\mu\nu}\,,\nn
\eea
where $V_e= -\alpha_e/|\bx|\,.$ We dropped the tag on the field, which now represents the wave-function of an electron in the background of a static Coulomb-like source with typical energy/momenta of order~$m_ev^2$. Notice the contribution from $c_V$ may be thought of as a local {\it renormalization} of the potential,
 \beq
 \delta V_e(\bx) = c_V \frac{e^2}{8m_e} \delta^3(\bx)\,. \label{dve}
\eeq

\subsection{The Lamb shift}

\begin{figure}[h!]
\centerline{{\includegraphics[width=0.3\textwidth]{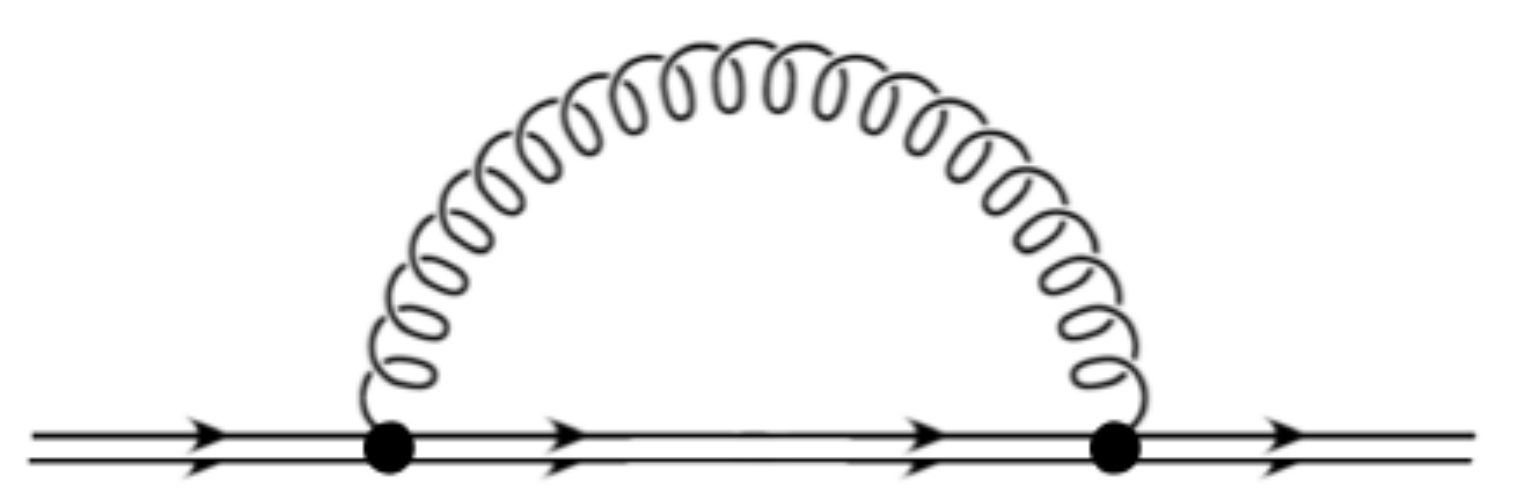}}}
\caption[1]{The one-loop correction in \eqref{green}. The double line represents the bound state, and the dots are the dipole-type coupling from \eqref{pnrqed}. A similar diagram --albeit at the classical level-- appears in NRGR (see~below).} \label{fig2}
\end{figure}
The calculation of the Lamb shift can be found in different textbooks, e.g. \cite{brown}. Here we derive it following the framework of the EFT approach NRQED. (The use of dim. reg. to regularize the divergences in the computation of the Lamb shift was also advocated in \cite{brown,brown2,lambeft}.)\vskip 4pt

The ultra-soft contribution to the $E_n$ level of the Hydrogen atom is represented in fig.~\ref{fig2}, and is given by a self-energy type diagram. The computation entails the two-point function
\beq
G(t,\bx) \equiv -i\langle 0|T(\psi(0)\psi(t,\bx))|0\rangle\,,
\eeq
which it is convenient to transform into Fourier space
\beq
\tilde G(\bx,E) = \int dt\, e^{iEt}\, G(t,\bx)\,.
\eeq
At leading order, introducing a complete set of states, we have
\beq
\tilde G_0(\bx,E) = \sum_{n,\ell} \frac{\psi_{n,\ell}(0)\psi^\dagger_{n,\ell}(\bx)}{E-E_n +i\epsilon}\,,
\eeq
where $E_n$ is the unperturbed energy level, with wave-functions $\psi_{n,\ell} \equiv \langle 0|\psi|n,\ell\rangle$, obeying
\beq
\hat H_0 \psi_{n,\ell} = E_n \psi_{n,\ell}\,,
\eeq
with \beq H_0 = \frac{\bp^2}{2m_e}+ V_e\,,\eeq the unperturbed non-relativistic Hamiltonian. The loop correction in fig.~\ref{fig2} contributes to the self-energy, $\Sigma(E)$, of the electron moving in a Coulomb background \cite{schwinger}. The one-loop diagram can be resumed as a Dyson series, leading to a correction to the Green's function, 
\beq
\label{green}
\left(E- \frac{\bp^2}{2m_e}-V_e  - \Sigma(E)\right)G(\bx,E) = 1\,,
\eeq 
and subsequently to the energy levels. Here $\bp^i$ is the momentum operator: $\bp^i = -i \boldsymbol{\nabla}^i$.\vskip 4pt The self-energy diagram can be computed in dim. reg. using the Feynman rules from \eqref{pnrqed}, and it reads\footnote{The (ultra-soft) photon propagator in Coulomb gauge is given by \beq
D^{ij}_{US}(k_0,\bk) = \frac{i}{k_0^2-\bk^2+i\epsilon}\left(\delta^{ij} - \frac{\bk^i\bk^j}{\bk^2}\right),\,\,\, D^{00}_{US}(k_0,\bk) = \frac{i}{\bk^2}\,.\eeq
The non-propagating component contributes a (tadpole) scaleless integral $\left(\int \tfrac{dk_0}{k_0}\right)$ that can be set to zero in dim. reg. \label{footDR}}
\beq
\Sigma(E) = -i\frac{e^2}{m_e^2} \int \frac{d^d k}{(2\pi)^d}\left(\delta^{ij} - \frac{\bk^i\bk^j}{\bk^2}\right)
\frac{1}{k_0^2-\bk^2+i\epsilon}\; \bp^i \; \frac{1}{H_0- E - k_0 + i\epsilon}\; \bp^j\,.
\eeq 
Using (see footnote 1)
\beq
 \int \frac{d^dk}{(2\pi)^d} 
\frac{1}{k_0^2-\bk^2+i\epsilon} \left(\delta^{ij} - \frac{\bk^i\bk^j}{\bk^2}\right) \frac{1}{\omega-k_0+ i\epsilon} = i \frac{\omega}{6\pi^2} \delta^{ij}\left(\frac{1}{\epsilon_{\rm UV}} + \frac{5}{6} - \log \frac{2\omega}{\mu}\right) \,, 
\eeq
we obtain,
\beq
\label{sigma}
\Sigma(E) = \frac{2\alpha_e}{3 \pi}\, \frac{\bp^i}{m_e} (H_0-E)\left(\frac{1}{\epsilon_{\rm UV}} + \frac{5}{6} - \log \frac{2(H_0-E)}{\mu}\right) \frac{\bp^i}{m_e}\,. \eeq 
Taking the limit $E \to E_n$, we find for the energy shift:
\bea
\label{eqDE}
(\delta E_{n,\ell})_{US} &=& \frac{2\alpha_e}{3\pi}\,\left[ e^2 \left(\frac{1}{\epsilon_{\rm UV}} + \frac{5}{6} \right) \frac{|\psi_{n,\ell}(\bx=0)|^2}{2m_e^2}  \right. \\  && - \left.  \sum_{m\neq n,\ell} \left\langle n,\ell\left|\frac{\bp}{m_e}\right|m,\ell\right\rangle^2 (E_m-E_n) \log \frac{2|E_n-E_m|}{\mu}\right]\nn
\eea
where we used \cite{brown}
\beq
\bp^i (H_0-E_n) \bp^i = \frac{1}{2}\boldsymbol{\nabla}^2 V_e =\frac{ e^2}{2}\delta^3(\bx)\,.
\eeq
To complete the relevant part of the calculation we need to add the (local) contribution from the short-distance modes in \eqref{dve}, proportional to the Wilson coefficient $c_V$ in \eqref{eqcd}, which yields
\beq
(\delta E_{n,\ell})_{c_V} = \langle n,\ell|\delta V_e|n,\ell \rangle = \frac{e^2}{8 m_e^2} c_V |\psi_{n,\ell}(\bx=0)|^2 = \frac{4\alpha_e^2}{3 m_e^2} \left( - 
\frac{1}{\epsilon_{\rm IR}}+ \log \frac{m_e}{\mu}\right) |\psi_{n,\ell}(\bx=0)|^2\,.
\eeq
Therefore, combining the two terms together we have
\bea
\delta E_{n,\ell} &=& (\delta E_{n,\ell})_{US}+ (\delta E_{n,\ell})_{c_V} + \cdots \label{lambS}\\
&=& \frac{2\alpha_e}{3\pi} \left[ \frac{5}{6} e^2 \frac{|\psi_{n,\ell}(\bx=0)|^2 }{2m_e^2} - \sum_{m\neq n,\ell} \left\langle n,\ell\left|\frac{\bp}{m_e}\right|m,\ell\right\rangle^2 (E_m-E_n) \log \frac{2|E_n-E_m|}{m_e}\right]+\cdots\nn \\ &&+ \frac{4\alpha_e^2}{3 m_e^2} \left(\frac{1}{\epsilon_{\rm UV}}-\frac{1}{\epsilon_{\rm IR}} \right) |\psi_{n,\ell}(\bx=0)|^2\,.
\nn\eea
Notice the anticipated link between IR and UV divergences. Provided we identify the IR/UV poles, these two singular terms drop out of the computation, as the factors of $\log \mu$ do. The relevant scale in the logarithm is replaced by $m_e$.  In the next sub-section we will describe how to properly implement the cancelation. The remaining terms are the celebrated correction in the Lamb shift at leading order, including Bethe logarithm and the numerical factor of $5/6$ \cite{Bethe,dyson,Viki,Kroll}. By power counting the (enhanced) logarithmic contribution, we find it scales as (recall $\alpha_e \sim v$)\footnote{One can actually think of two contributions, from $\log (E\, r_B)$ and (minus) $\log (m_e r_B)$, both scaling as $\log v$. In gravity, on the other hand, we only find a logarithm of the ratio between radiation and potential scales, at the desired order. Nevertheless, the basic steps are essentially the same in both cases.} 
\beq \delta E_{n,\ell} \simeq  \alpha_e v^2 m_e v^2 \log(m_ev^2/m_e) \sim m_e \alpha_e^5 \log \alpha_e\,.\eeq 
Notice that, if one treats the local contribution from $\delta V_e$ in \eqref{dve} independently, we would be misguided to remove the IR pole in \eqref{1} first, in order to arrive to a finite result. This, in turn, would introduce scheme-dependent ambiguities, since we could subtract from \eqref{1} either $1/\epsilon_{\rm IR}$ or $1/\epsilon_{\rm IR} + C$, with $C$ some unspecified dimensionless constant. Hence, after removing the UV divergence from the ultra-soft loop with an (independent) counter-term, we would need additional information to fix an undetermined contribution \cite{brown}
\beq
\delta V_e^{(C)} = C \frac{4\alpha_e^2}{3m_e^2} \delta^3(\bx)\,, 
\eeq
similarly to what occurs in the methodology in \cite{4pnjs,4pnjs2,4pndjs,4pndjs2,4pnbla,4pnbla2,4pndj}.
We discuss in what follows the steps which enable us to obtain an unambiguous result for the Lamb shift, regardless of the regularization scheme.

 \subsection{The zero-bin subtraction}\label{zero}

We must implement a procedure in which modes other than the ultra-soft never leave the realm pertinent to the bound state, henceforth avoiding IR divergences. This is known as the zero-bin subtraction~\cite{zero}. As an example, let us consider any one-loop graph in NRQED with contributions from different regions. Let us concentrate only on the propagating degrees of freedom, namely soft and ultra-soft modes. The soft part of the graph may have UV and IR divergences, 
\beq
I_S = \frac{A_S}{\epsilon_{\rm UV}} + \frac{B_S}{\epsilon_{\rm IR}} + f_S(q,\mu)\,,
\eeq
with $q \sim m_e v$. The UV divergence is removed by a counter-term as usual, therefore, without loss of generality, we set $A_S=0$. On the other hand, for the ultra-soft part,
\beq
I_{US} = \frac{A_{US}}{\epsilon_{\rm UV}} + \frac{B_{US}}{\epsilon_{\rm IR}} + f_{US}(E,\mu)\,,
\eeq
with $E \sim m_e v^2$. The IR divergences in the ultra-soft calculation would match into the IR singularities of the full theory, if any, in the quantity at hand. Let us assume the observable is IR-safe in QED, and therefore $B_{US} =0$. Since the method of regions is designed to reproduce the full theory computation in terms of relevant zones, we must have \cite{Beneke,IraTasi}
\beq
I_{\rm full} = I_S + I_{US} + I_{\rm hard}\,,
\eeq
where the `hard' part corresponds to modes with $k \sim m_e$. This is the contribution which matches into Wilson coefficients, as a series of local terms.\footnote{The method of regions and dim. reg. go hand-by-hand, enforcing that contributions from momenta $k \gg m_e$ can be ignored, since they turn into a scaleless integral.}\vskip 4pt
In general, we will find $B_S = - A_{US}$, which will be ultimately related to the cancelation of spurious divergences due to the splitting into regions. Therefore, adding the soft and ultra-soft contributions together,
\beq
I_S+I_{US} =   f_S(q,\mu) + f_{US}(E,\mu) + B_S\left( \frac{1}{\epsilon_{\rm IR}} -\frac{1}{\epsilon_{\rm UV}} \right)\,.
\eeq
The role of the zero-bin subtraction is to remove from $I_S$ the IR singularity. In other words, we replace \beq I_S \to I_S - I_{\rm zero\text{-}bin},\eeq where $I_{\rm zero\text{-}bin}$ corresponds to an asymptotic expansion of the soft integral around the region responsible for the IR poles. This procedure removes the double-counting induced by the overlap between the IR sensitive part of the $I_S$ integral and the contribution from $I_{US}$.\vskip 4pt The zero-bin part may involve a scaleless integral, which in dim. reg. are usually set to zero. That is the case because they entail a {\it cancelation} between IR and UV poles. However, when IR divergences are present, scaleless integral require some extra care \cite{IraTasi}. In dim. reg., the zero-bin will often take the form, 
\beq
I_{\rm zero\text{-}bin} = B_S \left( \frac{1}{\epsilon_{\rm IR}} - \frac{1}{\epsilon_{\rm UV}}\right)+\, \rm{finite}\,,
\eeq
such that
\beq
I_S-I_{\rm zero\text{-}bin} + I_{US} =  f_S(q,\mu) + f_{US}(E,\mu) \,,
\eeq
See \cite{zero} for more details.\vskip 4pt

Returning to the case at hand, there are a few subtleties regarding the IR divergence in~\eqref{1}. In~principle, the IR pole entered in the matching into NRQED.\footnote{Technically speaking, QED is first matched into HQET by integrating out $m_e$. The same happens in the gravitational case, with the finite size scale identified with the hard modes.} However, an effective theory is constructed such that all the long-distance physics from the full theory is recovered. Hence, the IR divergence in \eqref{1}, which trickled into $c_V$ in \eqref{eqcd}, should be matched to a similar IR singularity in the effective theory \cite{IraTasi}. The IR pole in the EFT side, however, is subtle, since it arises from scaleless integrals which are often ignored \cite{manohar}.\footnote{Notice that, while adding a scaleless integral from the EFT side may cancel the IR poles on both sides of the matching condition, it also leaves behind a UV divergent term, as in the zero-bin prescription. The latter would likewise cancel out against the UV divergence in the ultra-soft loop.} At the end of the day, this procedure (keeping scaleless integrals in the long-distance theory) is entirely equivalent to performing a zero-bin subtraction from $I_{\rm hard}$, removing unwanted soft(er) modes prior to performing the matching. The advantage of implementing the zero-bin prescription is that it enables us to set to zero other scaleless integrals (for example the contribution from $A^0_{US}$ in the calculation of the Lamb shift, see footnote \ref{footDR}), since all quantities are then IR-safe. (Moreover, the zero-bin subtraction is independent of the regularization scheme.)
\vskip 4pt 
Let us return to the form factor in \eqref{1}. If we denote as $(p , p-q)$ the incoming and outgoing momenta respectively, the vertex correction entails 
\beq
I_{\rm vertex}= -ie^2 p\cdot(p-q) \int \frac{d^dk}{(2\pi)^d}\frac{1}{k^2+i\epsilon} \frac{1}{(p-k)^2-m_e^2+i\epsilon}  \frac{1}{(p-q-k)^2-m_e^2+i\epsilon}\,.
\eeq 
The part of the integral with $k \sim m_e v$ is reproduced by the soft modes in NRQED, and likewise for the ultra-soft modes. On the other hand, the contribution from the hard region, which matches into Wilson coefficients, is given by modes with $k \sim m_e$. At leading order in $q^2/m^2_e$ we have,
\beq
I_{\rm hard} = -ie^2 m_e^2 \int \frac{d^dk}{(2\pi)^d}\frac{1}{k^2+i\epsilon} \left(\frac{1}{k^2-p\cdot k +i\epsilon}\right)^2 + {\cal O}(q^2/m_e^2)\,.
\eeq 
This integral clearly has an IR divergence, and the result reads
\beq
I_{\rm hard} = \frac{e^2}{8\pi^2} \left(\frac{1}{\epsilon_{\rm IR} }+ \log \mu/m_e\right)+{\cal O}(q^2/m_e^2)\,.
\eeq
The IR pole, however, appears from the region, $k \ll m_e$, which does not belong to $I_{\rm hard}$. Therefore, we need to perform the (zero-bin) subtraction
\beq
I_{\rm zero\text{-}bin} = -ie^2  \int \frac{d^dk}{(2\pi)^d}\frac{1}{k^2+i\epsilon} \left(\frac{1}{v \cdot k +i\epsilon}\right)^2\,,
\eeq
where we used $p^\mu = m_e v^\mu$, and $p^2=m_e^2$. This integral is easy to calculate in the rest frame, with $v^\mu=(1,0,0,0)$, yielding
\beq
I_{\rm zero\text{-}bin}  = \frac{e^2}{8\pi^2}\left(\frac{1}{\epsilon_{\rm IR}}- \frac{1}{\epsilon_{\rm UV}}\right)\,,
\eeq
such that
\beq
I_{\rm hard} - I_{\rm zero\text{-}bin} = \frac{e^2}{8\pi^2} \left(\frac{1}{\epsilon_{\rm UV} }+ \log \mu/m_e\right)\,.
\eeq
Iterating this procedure in all the IR sensitive terms transforms the IR pole in \eqref{eqcd} into a UV singularity, 
\beq
\label{eqcdz}
c_V \xrightarrow{\rm zero\text{-}bin}  1 + \frac{8}{3} \frac{\alpha_e(\mu)}{\pi} \left[ - \frac{1}{\epsilon_{\rm UV}}+  \log m_e/\mu\right]\,.
\eeq 
Following our computation of the Lamb shift, this UV pole now readily cancels against the UV divergence arising in the ultra-soft loop correction, see \eqref{lambS}, unfolding the ambiguity-free final result. The same would have happened had we used any other regularization scheme. 

\subsection{The renormalization group} \label{sec:rg}

In the previous calculation within NRQED we ended up without divergences, but also the factors of $\mu$ are gone after using \eqref{eqcdz}. However, we could have approached the problem differently --from the bottom up-- by computing directly in the ultra-soft effective theory. While the matching condition determines the value of the parameters in the effective theory (at a matching scale), the from of the effective Lagrangian can be constructed using the low-energy symmetries and degrees of freedom \cite{eftCL}. There is (at least for our purposes) only one Wilson coefficient, $c_V$, in the long-distance theory. The computation of the shift in the energy levels follows from the ultra-soft loop, which is UV divergent. From the point of view of the ultra-soft theory we can then use a counter-term to renormalize the divergence. Hence, the UV pole may be removed via
\beq
c_V^{\rm c.t.} = -\frac{8\alpha_e }{3\pi} \frac{1}{\epsilon_{\rm UV}}\,,
\eeq 
or in terms of the local potential (see \eqref{dve})
\beq
\delta V_e^{\rm c.t.} = -\frac{4\alpha_e^2}{3 m_e^2} \frac{1}{\epsilon_{\rm UV}}\,\delta^3(\bx)\label{vect}\,.
\eeq 
Putting the pieces together, we find
\bea
\delta E_{n,\ell} &=&  \left[ \left\{\frac{2\alpha_e}{3\pi}\left(\frac{5}{6} + \log \frac{\mu}{m_e} \right) + \frac{c^{\rm ren}_V(\mu)}{4} \right\}e^2\frac{|\psi_{n,\ell}(\bx=0)|^2 }{2m_e^2} \right. - \nn \\ & & \left. \frac{2\alpha_e}{3\pi} \sum_{m\neq n,\ell} \left\langle n,\ell\left|\frac{\bp}{m_e}\right|m,\ell\right\rangle^2 (E_m-E_n) \log \frac{2|E_n-E_m|}{m_e}\right]+\cdots\,. \label{eqDE2}
\eea
Notice two important differences. First of all, the appearance of a renormalized parameter, $c^{\rm ren}_V(\mu)$, and the $\log \mu$. The binding energy is obviously $\mu$-independent, and therefore one can obtain a renormalization group equation,
\beq
\mu \frac{d}{d\mu} \delta E_{n,\ell} = 0 \to \mu \frac{d}{d\mu} c_V^{\rm ren}(\mu) = -\frac{8\alpha_e(\mu)}{3\pi}\,,
\eeq
or, in other words,
\beq
\mu \frac{d}{d\mu} \delta V^{\rm ren}_e(\bx,\mu) = -\frac{4\alpha_e^2}{3 m_e^2} \delta^3(\bx)\label{verg}\,.
\eeq
By solving this equation we find,\footnote{To be consistent we should match pNRQED into NRQED at $\mu_0 \sim m_e v$. However, since the zero-bin subtraction removes the double counting, we can {\it pull-up} the matching condition to $\mu_0 \sim m_e$. (See fig.~1 in \cite{zero}, also \cite{vrg,ManSt,hoang} for the implementation of the `velocity renormalization group'  in `vNRQED', which is better suited to handle the $\log v$'s to all orders in $\alpha_e$ (and $\alpha_s$) in one go, from $m_e$ to $m_ev^2$.)} 
\beq
c_V^{\rm ren}(\mu) = c_V^{\rm ren}(m_e) - \frac{8\alpha_e}{3\pi} \log \frac{\mu}{m_e}\,,
\eeq
and likewise (in momentum space)
\beq
\delta V^{\rm ren}_e(\bp,\mu) = \delta V^{\rm ren}_e(\bp,m_e) - \frac{4\alpha_e^2}{3 m_e^2}  \log \frac{\mu}{m_e}\,.
\eeq
The utility of this expression is clear. First of all, let us re-write \eqref{eqDE2} as
\beq
\delta E_{n,\ell} =  \left\langle n,\ell\left|\delta V^{\rm ren}_e(\bx,\mu) \right| n,\ell \right\rangle  -  \frac{2\alpha_e}{3\pi} \sum_{m\neq n,\ell} \left\langle n,\ell\left|\frac{\bp}{m_e}\right|m,\ell\right\rangle^2 (E_m-E_n) \log \frac{2|E_n-E_m|}{\mu}+\cdots\,.\label{eqDE3}
\eeq
If we now take $\mu \sim m_e v^2$, the second term in \eqref{eqDE3} becomes subdominant (since $\Delta E/(m_ev^2) \sim 1$). Hence, we directly obtain the logarithmic Lamb shift from the renormalization group equation (recall $\alpha_e \sim v$)
\begin{align}
\delta E_{n,\ell} &=  \left\langle n,\ell\left|\delta V^{\rm ren}_e(\bx,\mu= m_e v^2) \right| n,\ell \right\rangle  +\cdots = -\frac{4\alpha_e^2}{3m_e^2} |\psi_{n,\ell}(\bx=0)|^2 \log v^2 +\cdots 
\\& = - \frac{8}{3\pi} \frac{\delta_{\ell 0}}{n^3}\, m_e \alpha_e^5 \log \alpha_e + \cdots\,,\nn
\end{align}
where (only the $\ell=0$ states have support at $\bx=0$)
\beq
|\psi_{n,\ell}(\bx=0)|^2=  \frac{\alpha^3_e m^3_e}{\pi n^3}\delta_{\ell0}\,,
\eeq
for the Hydrogen atom.
In this manner we unambiguously obtain Bethe logarithm directly from the long-distance effective theory. This is similar to what we find in the gravitational case, which we discuss~next.

\section{The (classical) binding energy in gravity}

The two-body problem in gravity, needless to say, is classical in nature, whereas the Lamb shift in QED is rooted in quantum effects. Moreover, gravity is in spirit more closely related to the strong interaction, and NRQCD, where the potential and ultra-soft gauge fields can couple not only to fermions but also to each other \cite{IraTasi}. Nonetheless, similarities arise between the two EFT approaches. In NRGR, as in NRQED, the IR divergence in the near region is also linked to a UV pole in the far zone. The latter follows from a {\it conservative} radiative effect, namely the tail contribution to the radiation-reaction force \cite{nltail}. Moreover, akin to the implementation in electrodynamics, the IR divergences can be removed using the zero-bin subtraction, paving the way to ambiguity-free results \cite{comp}. To complete the analogy, in what follows we rederive the logarithmic correction to the binding potential for binary black holes, which bears a close resemblance with our derivation of Bethe logarithm for the Hydrogen atom.

\begin{figure}[t!]
\centerline{{\includegraphics[width=0.55\textwidth]{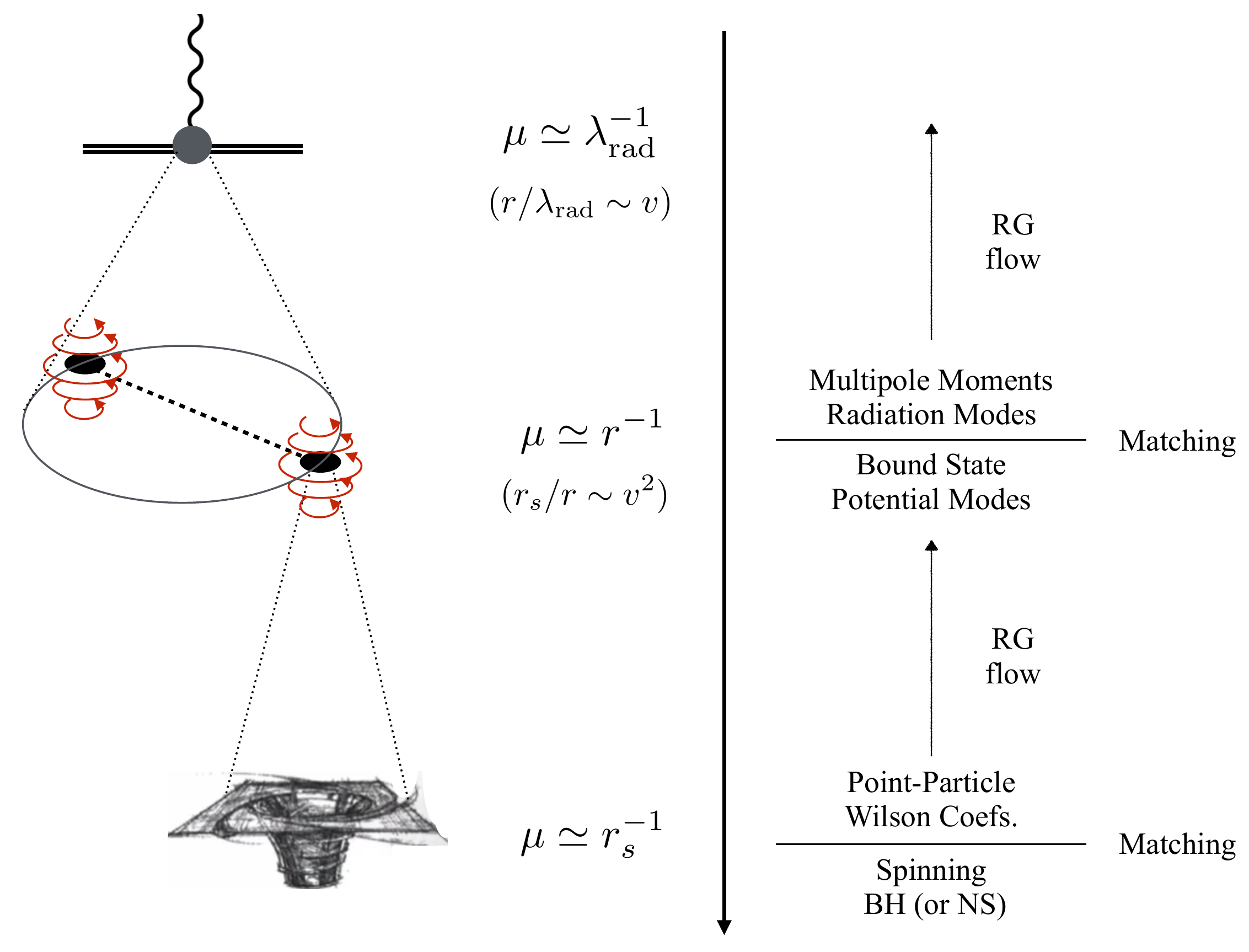}}}
\caption[1]{The EFT approach to the binary inspiral problem. See \cite{review} for a thorough review.} \label{spin2}
\end{figure}
\subsection{The EFT framework: NRGR}
The relevant scales for the binary inspiral problem are, the size of the compact object, $r_s$, the separation,~$r$, and the typical wavelength of the emitted radiation, $\lambda_{\rm rad} \sim r/v$. For a bound state we also have $r_s/r \sim v^2$, and therefore
\beq
r_s \ll r \ll \lambda_{\rm rad}\,,
\eeq
in the PN regime, $v \ll 1$. Therefore, after the hard scale, $r_s$, is integrated out we encounter two relevant regions for the binary problem (recall soft modes are not present in classical computations). Namely, the --off-shell-- potential,
\beq
(p^0_{\rm pot},\bp_{\rm pot}) \sim (v/r,1/r)\,, \label{pot}
\eeq
and --on-shell-- radiation (or ultra-soft) modes,
\beq
(p^0_{\rm rad},\bp_{\rm rad}) \sim (v/r,v/r)\,.\label{rad}
\eeq
The NRGR action takes the form ($L=i_i\ldots i_\ell$)\footnote{As in electrodynamics, the expression in \eqref{nrgr} applies more generally to the dynamics of an extended objects in a long-wavelength background, prior to considering a two-body bound state \cite{review}.}
\bea
 \nn
 S_{\rm NRGR} [x^{(p)}_{\rm cm}(\tau),h_{\mu\nu}] &=& \sum_p \int d\tau_p \left[ -M_{(p)}(\tau) - \frac{1}{2} \omega_{\mu \, ab} S_{(p)}^{ab}(\tau) u^\mu(\tau) \right. \\  &+& \left.  \sum_{\ell=2} \left( \frac{1}{\ell!} I_{{\rm src}(p)}^L(\tau) \nabla_{L-2} E_{i_{\ell-1}i_\ell}- \frac{2\ell}{(2\ell+1)!}J_{{\rm src}(p)}^L(\tau) \nabla_{L-2} B_{i_{\ell-1}i_\ell}\right)\right]\,,\label{nrgr}
\eea
where $x^{(p)}_{\rm cm}(\tau)$ is the center-of-mass worldline of the bodies, $\omega_{\mu\, ab}$ are the Ricci coefficients, and $E_{ij}$, $B_{ij}$ are the electric and magnetic components of the Weyl tensor. The metric perturbation, $h_{\mu\nu} = g_{\mu\nu} - \eta_{\mu\nu}$, has support on modes longer than the hard scale, and it includes both potential and radiation modes. The monopole, $M$, represents the mass, $S_{ab}$ is the spin tensor, and the $I^L_{\rm src},J^L_{\rm src}$ are the permanent mass- and current-type source multipole moments, of the compact objects \cite{review}.\footnote{For instance, for a spinning body $I_{\rm src}^{ij} = \frac{1}{2} C_{ES^2} S^{ik}S^j_k$ \cite{nrgrs,prl,nrgrs1s2,nrgrs1s1,review}. We must also incorporate response terms, e.g. to the background field induced by the companion, $I^{ij}_{\cal R} = C_E E^{ij} +\cdots$, and likewise for the magnetic components. The $C_{E,B}$ coefficients are known as Love~numbers, encoding the information regarding the internal degrees of freedom of the compact bodies. (Surprisingly, all the Love numbers vanish for black holes in $d=4$, which opens up a unique opportunity to test the {\it shape} of spacetime in the forthcoming era of precision gravity \cite{love,cardoso,iragrg}.)}

\vskip 4pt The EFT for at the radiation scale is constructed similarly to pNRQED (although at the non-linear level the structure resembles pNRQCD instead), by integrating out the potential modes \cite{review}. Unlike~QED, all the calculations remain at the classical level, involving a series of iterations of Green's functions convoluted with external sources. Because of the symmetries of the long-distance theory, i.e. general relativity, the effective action in the radiation sector is exactly the same as in \eqref{nrgr}, but only radiation fields are present. The bodies are replaced by a single worldline at the center-of-mass of the binary, and the Wilson coefficients are now associated with the two-body system. For example, $M$ is the (Bondi) binding energy of the bound state, and $(I^L_{\rm src},J^L_{\rm src})$ are the corresponding source multipole moments. In principle, the power loss is obtained in terms of their time derivatives, using the equations of motion which follow from the gravitational binding potential \cite{review}. See 
fig.~\ref{spin2} for a schematic representation of the relevant scales in NRGR. There is yet one other important contribution to be considered, namely the tail effect, or the scattering of the outgoing radiation off of the Newtonian potential produced by the whole binary. This is responsible for the rich structure of the radiation theory \cite{andirad,andirad3,nltail}.

\subsection{The tail effect}
\begin{figure}[t!]
\centerline{{\includegraphics[width=0.15\textwidth]{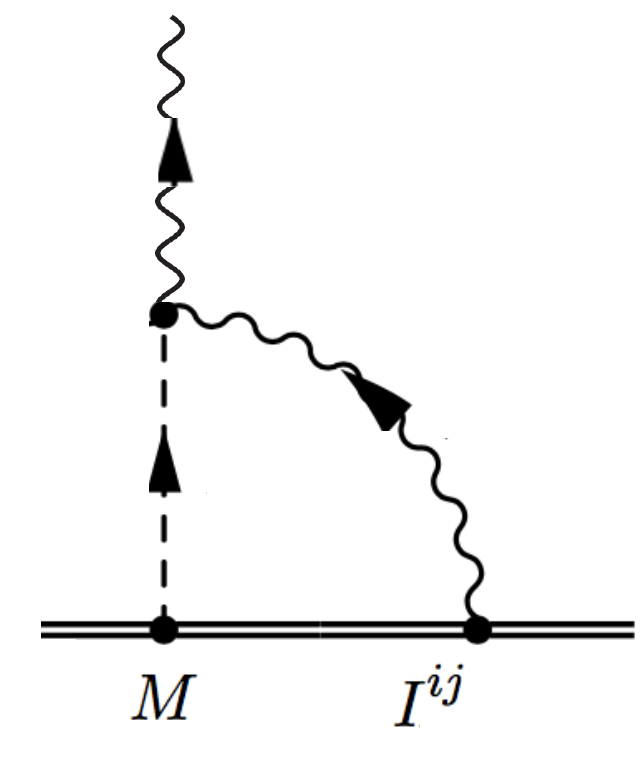}}}
\caption[1]{The tail contribution to the radiative quadrupole moment. Only the lines with an arrow propagate. The double-line represents the two-body system, treated as an external non-propagating source.} \label{amp}
\end{figure}
\begin{figure}[t!]
\centerline{{\includegraphics[width=0.25\textwidth]{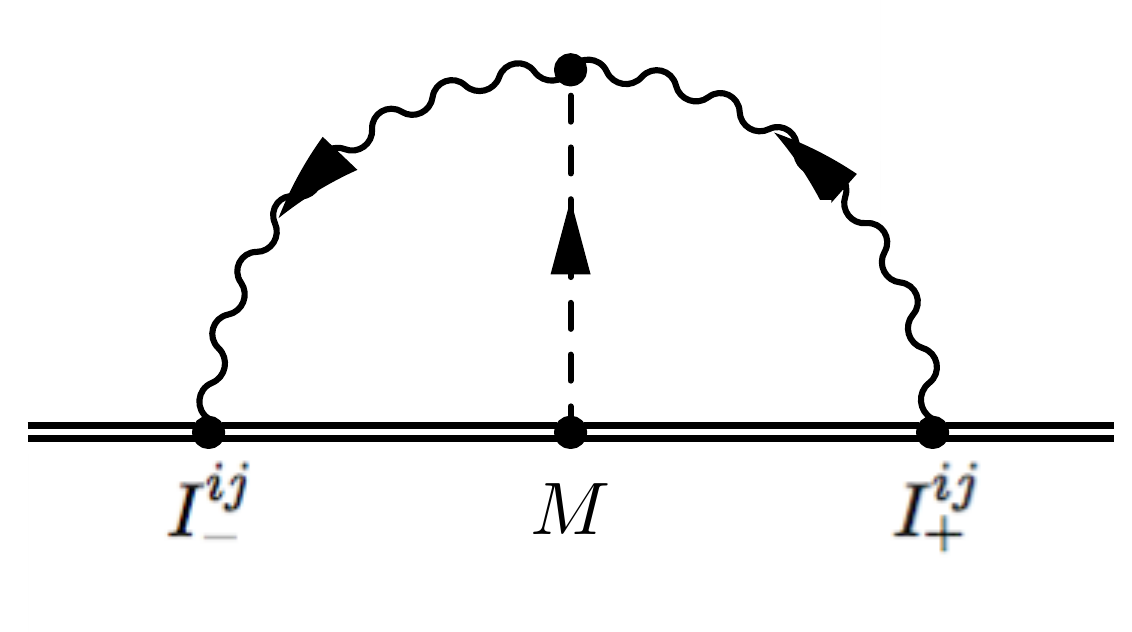}}}
\caption[1]{The tail contribution to radiation-reaction. The $(+,-)$ labels are associated to the `in-in' formalism, required to properly compute retardation effects. The wavy line is a radiation mode $p^\mu \sim \lambda_{\rm rad}^{-1}$, whereas the dashed line corresponds to a potential mode with $\bq \sim \lambda_{\rm rad}^{-1}$. See \cite{nltail} for more details.} \label{nltail}
\end{figure}
\begin{figure}[t!]
\centerline{{\includegraphics[width=0.65\textwidth]{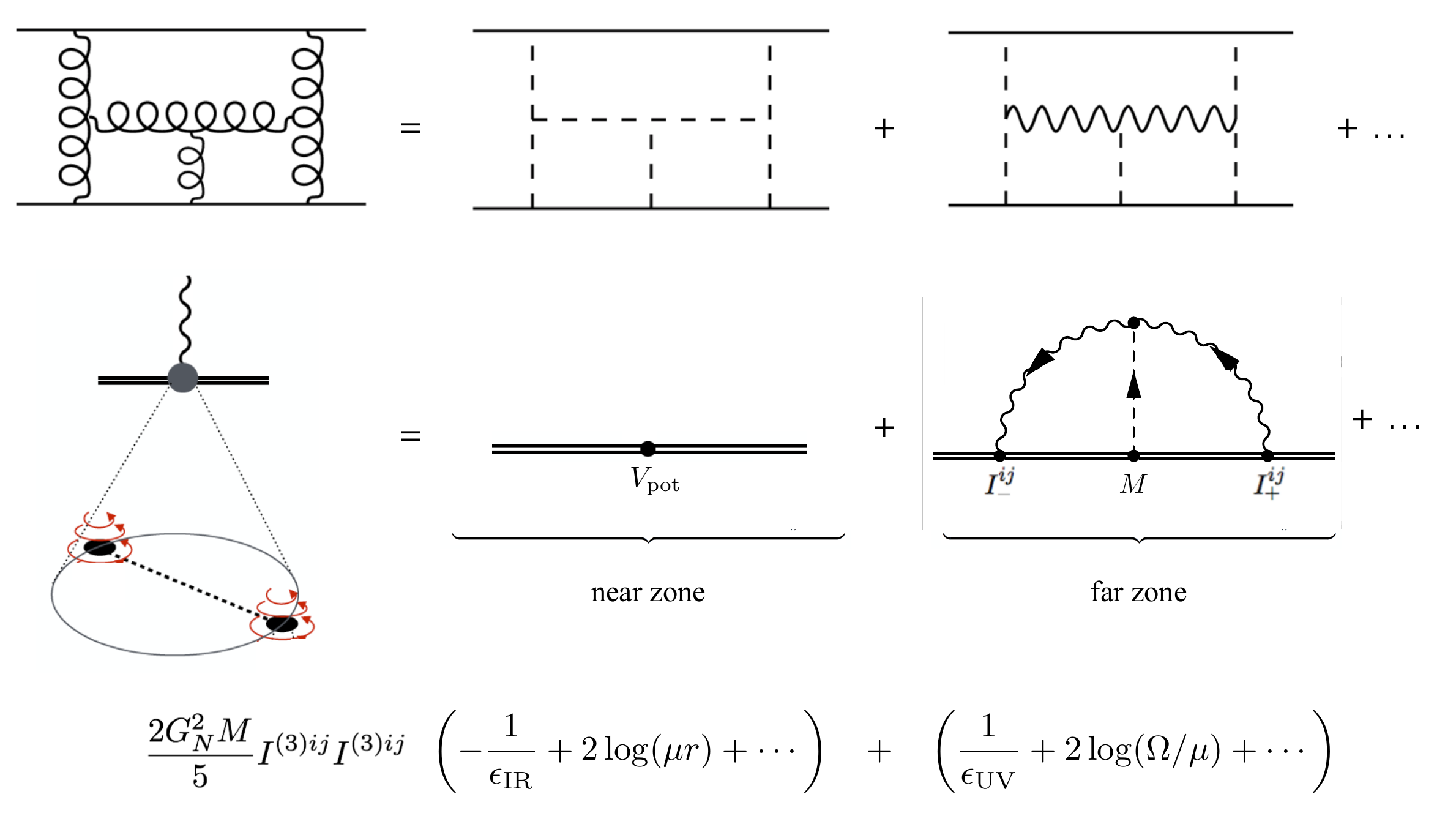}}}
\caption[1]{The full theory computation in general relativity (curly propagators) is split into regions in the EFT formalism: potential (dashed) and radiation (wavy) modes. The subsequent IR/UV divergences appear from the splitting into near and far zones \cite{nltail}. The calculations are similar to the derivation of the (quantum) Lamb shift described here, except that the gravitational case involves non-linear couplings (and is fully classical). The logarithmic contribution, scaling as $ M v^8\log v$, resembles Bethe logarithm in electrodynamics.\label{glue}} 
\end{figure}

The interaction of the binary's gravitational potential with the outgoing radiation modifies the total emitted power. In practice, the source moments, $I^{L}_{\rm src}$, which enter in the effective action in \eqref{nrgr}, turn into radiative multipoles, $I^{L}_{\rm rad}$, in the computation of the radiated power \cite{blanchet}. For example, the radiative quadrupole is obtained by computing the Feynman graph in fig.~\ref{amp}, which follows from the interaction between the quadrupole, $I_{\rm src}^{ij}$, and the monopole, $M$. The calculation is straightforward, and one obtains a correction of the form \cite{andirad,amps, tail3,tail3n},
\beq
I^{ij}_{\rm rad}(\omega) = I^{ij}_{\rm src}(\omega)\left[1+ GM \omega
 \left({\rm sign}(\omega) \pi + i\left[ \frac{2}{\epsilon_{\rm IR} }+ \log \omega^2/\mu^2 +{\rm finite}\right]\right)\right]\,,
\eeq
which features an IR divergence. It is easy to see all the IR poles cancel out in the radiated power, since they add up to an overall phase \cite{andirad}. (This type of IR divergence is thus intimately related to the soft factors in QED  \cite{weinberg}.) However, similarly to what occurred for the Lamb shift, the contribution from the tail effect to radiation-reaction, and in particular its {\it conservative} part, features instead a UV divergence, see fig.~\ref{nltail},\footnote{There is also a  $i \pi\text{sign}(\omega)$ in the computation which accounts for the radiative part of the tail contribution, see \cite{nltail}.} 
\beq
\int dt\, V_{\rm tail}(\mu) =    \, \frac{G_N^2 M}{5} \int \frac{d\omega}{2 \pi} \,  \omega^6 \, I_{\rm src}^{ij}(-\omega) I_{\rm src}^{ij}(\omega) \left[  \frac{1}{ \epsilon_{\rm UV}} + \log \frac{\omega^2}{\mu^2}+ {\rm finite}\right] .
\label{eq:RRnl}
\eeq
(We drop the `src' label below since all the multipole moments in what follows refer to the source.) The~term in \eqref{eq:RRnl} is the equivalent to \eqref{eqDE} in the derivation of the Lamb shift. By the same token, the IR divergence in the NRGR potential from the near region (which enters as a local term in the radiation theory)  is the analogous~to~the one in \eqref{dve}, through \eqref{eqcd}. All we need is to show that the coefficients of the poles (and~the~$\log\mu^2$) match, as they do in NRQED.\vskip 4pt While the computation of the 4PN gravitational potential within the EFT approach is still undergoing \cite{nrgr4pn1,nrgr4pn2}, we expect to find the following structure in the near region \cite{nltail,comp}
\beq
\label{eq:V4pn1}
\int dt \, V_{\rm pot} (\mu) = -\frac{G_N^2 M}{5} \int \frac{d\omega}{2\pi} \omega^6 I^{ij}(-\omega) I^{ij}(\omega) \left(\frac{1}{\epsilon_{\rm IR}} - 2\log (\mu r)\right) + {\rm local/finite} \,.
\eeq
Hence, adding both contributions together, and restricting to a circular orbit (for which $\omega \simeq 2v/r$), we would get \cite{nltail} (see fig.~\ref{glue})
\beq
V_{\rm full} = V_{\rm pot} + V_{\rm tail} =  \frac{2G_N^2 M}{5}  I^{ij(3)}(t) I^{ij(3)}(t)\left[ \log v+ \frac{1}{2} \left(\frac{1}{\epsilon_{\rm UV}}-\frac{1}{\epsilon_{\rm IR}}\right)\right] + {\rm finite}\,.\label{vtot}
\eeq
The uperscript $(n)$ represents the $n$-th time derivative. In \cite{comp} we elaborate on the zero-bin prescription to deal with the divergences in \eqref{vtot}, which are the source of ambiguities in the regularization schemes implemented in \cite{4pnjs,4pnjs2,4pndjs,4pndjs2,4pnbla,4pnbla2,4pndj}. The logarithmic correction, on the other hand, is universal \cite{nltail}. The latter may be obtained unambiguously without the need of any matching condition, as we show next.  
\vskip 4pt

\subsection{The renormalization group} \label{sec:rg2}

As we did for the Lamb shift, let us proceed from the bottom up, where the gravitational potential from the near zone becomes a matching coefficient in the far zone. Therefore, as before (see e.g. \eqref{vect}), we split the local contribution from the near region into a renormalized part and a counter-term. The latter is chosen to renormalize the --conservative-- contribution from the tail effect \cite{nltail}
\beq
V_{\rm c.t.} =  - \frac{G_N^2 M}{5} I^{ij(3)}(t) I^{ij(3)}(t) \frac{1}{ \epsilon_{\rm UV}},
\eeq
so that we end up with a full gravitational potential of the form
\beq
\label{vtot2}
V_{\rm full} =  V_{\rm ren}(\mu) + \frac{G_N^2 M}{5} \int \frac{d\omega}{2 \pi} \,  \omega^6 \, I^{ij}(-\omega) I^{ij}(\omega) \left[ \log \frac{\omega^2}{\mu^2}+ {\rm finite}\right] .
\eeq
This expression is similar to \eqref{eqDE3}. Hence, by demanding the $\mu$-independence of the (physical) gravitational potential \cite{nltail} we find,
\beq
\mu\frac{d}{d\mu} V_{\rm full} = 0 \to \mu\frac{d}{d\mu} V_{\rm ren}(\mu) = \frac{2G_N^2 M}{5} I^{ij(3)}(t) I^{ij(3)}(t)\,,
\eeq
which is the equivalent of \eqref{verg}.
Once again, considering a circular orbit and choosing $\mu \sim v/r$, the renormalization group equation carries the information about the logarithmic contribution,
\beq
V_{\rm full}^{\rm \log} =  \frac{2G_N^2 M}{5}  I^{ij(3)}(t) I^{ij(3)}(t) \log v\,,
\eeq
reproducing \eqref{vtot}. From here, following the step described in \cite{nltail}, we derived the logarithm entering in the (conserved) binding energy at 4PN order,
\beq
\label{elog}
E_{\rm log} = -2 G_N^2 M  \big\langle I^{ij(3)}(t) I^{ij(3)}(t)\big\rangle \log v\,,
\eeq
which agrees with the result in \cite{ALTlogx}.

\section{Concluding remarks}

In this paper we studied the Lamb shift using NRQED, illustrating an ambiguity-free derivation of the binding energy within an EFT framework. The parallel with the gravitational case was already emphasized in \cite{4pnjs,4pnjs2}, quote: ``It is worth pointing out that also the Lamb shift calculation of Ref.~\cite{brown} shows up an undefined constant in the IR sector, which gets fixed by some dimensional matching."\footnote{The prescription in \cite{brown} is akin to a cancelation between IR and UV poles in dim. reg. (also advocated in \cite{nltail}). This is correct, yet conceptually distinct to the zero-bin subtraction. The latter may be applied to any regularization scheme (e.g. momentum cut-off \cite{zero}), including those used~in~\cite{4pnjs,4pnjs2,4pndjs,4pndjs2,4pnbla,4pnbla2,4pndj}, whereas the procedure in \cite{brown} only applies in dim. reg.}
 Indeed, an IR singularity appears in the near zone calculations in NRQED, resembling the situation in gravity. Likewise, a UV pole arises from an ultra-soft loop in the far region, echoing the calculation of the (conservative part of) the tail effect in NRGR \cite{nltail}. Yet, as we showed, the IR/UV divergences in the Lamb shift can be removed without the need to introduce ambiguities. The procedure is implemented for NRGR in \cite{comp}. We also rederived the renormalization group equations from which we reproduce both logarithmic contributions, to the --quantum-- shift in the energy levels of the Hydrogen atom and the --classical--  gravitational binding potential for binary black~holes. 

 \section*{Acknowledgments}
I am grateful to Ira Rothstein for enlightening conversations and Aneesh Manohar for comments on~a~draft. I~thank the participants of the workshop `Analytic Methods in General Relativity' held at ICTP-SAIFR,\footnote{\small \url{http://www.ictp-saifr.org/gr2016}} (supported by the S\~ao Paulo Research Foundation (FAPESP) grant 2016/01343-7) for very fruitful discussions, in particular to Luc Blanchet, Guillaume Faye, and Gerhard Sch\"afer. I thank Gerhard for bringing to my attention Ref. \cite{brown}, which prompted the ambiguity-free derivation of the Lamb shift presented here. I also thank the theory group at DESY (Hamburg) for hospitality while this paper was being completed. This work was supported by the Simons Foundation and FAPESP Young Investigator Awards, grants 2014/25212-3 and 2014/10748-5.

\bibliographystyle{utphys}
\bibliography{RefLamb}

\end{document}